\documentclass[%
 reprint,
 amsmath,
 amssymb,
 aps,
 superscriptaddress,
 prl,
]{revtex4-2}

\usepackage{xcolor}
\usepackage{physics}
\usepackage{dcolumn}
\usepackage{bm}
\usepackage{color}
\usepackage{textcomp}
\usepackage{hyperref}
\hypersetup{
  colorlinks=true,
  allcolors=blue
}
\usepackage{graphicx}
\usepackage{makecell}
\usepackage{here}

\newcommand{\reviseAdd}[1]{\textcolor{black}{#1}}
\newcommand{\tcolorOrange}[1]{\textcolor{black}{#1}}

\newcommand{\beginsupplement}{%
  \setcounter{table}{0}%
  \setcounter{figure}{0}%
  \setcounter{equation}{0}%
  \setcounter{section}{0}%
  \setcounter{enumiv}{0}
  \renewcommand{\thetable}{S\arabic{table}}%
  \renewcommand{\thefigure}{S\arabic{figure}}%
  \renewcommand{\theequation}{S\arabic{equation}}%
  \renewcommand{\thesection}{S\arabic{section}}%
}

\begin{document}

\preprint{APS/123-QED}

\title{
Selective Excitation of Superconducting Qubits \\with a Shared Control Line through Pulse Shaping
}

\author{R.\,Matsuda}
\thanks{These authors contributed equally to this work.}
\affiliation{Graduate School of Engineering Science, \tcolorOrange{The University of Osaka}, 1-3 Machikaneyama, Toyonaka, Osaka, 560-8531, Japan}
\affiliation{Center for Quantum Information and Quantum Biology, \tcolorOrange{The University of Osaka}, 1-2 Machikaneyama, Toyonaka, Osaka, 560-0043, Japan}
\email{u657789f@ecs.osaka-u.ac.jp}

\author{R.\,Ohira}
\thanks{These authors contributed equally to this work.}
\affiliation{\tcolorOrange{QuEL, Inc., Hachioji ON Building 5F, 4-7-14 Myojincho, Hachioji, Tokyo, Japan}}
\email{ohira@quel-inc.com}

\author{T.\,Sumida}
\affiliation{\tcolorOrange{QuEL, Inc., Hachioji ON Building 5F, 4-7-14 Myojincho, Hachioji, Tokyo, Japan}}

\author{H.\,Shiomi}
\affiliation{Center for Quantum Information and Quantum Biology, \tcolorOrange{The University of Osaka}, 1-2 Machikaneyama, Toyonaka, Osaka, 560-0043, Japan}
\affiliation{\tcolorOrange{QuEL, Inc., Hachioji ON Building 5F, 4-7-14 Myojincho, Hachioji, Tokyo, Japan}}

\author{A.\,Machino}
\affiliation{Graduate School of Engineering Science, \tcolorOrange{The University of Osaka}, 1-3 Machikaneyama, Toyonaka, Osaka, 560-8531, Japan}
\affiliation{Center for Quantum Information and Quantum Biology, \tcolorOrange{The University of Osaka}, 1-2 Machikaneyama, Toyonaka, Osaka, 560-0043,  Japan}

\author{S.\,Morisaka}
\affiliation{Center for Quantum Information and Quantum Biology, \tcolorOrange{The University of Osaka}, 1-2 Machikaneyama, Toyonaka, Osaka, 560-0043, Japan}
\affiliation{\tcolorOrange{QuEL, Inc., Hachioji ON Building 5F, 4-7-14 Myojincho, Hachioji, Tokyo, Japan}}

\author{K.\,Koike}
\affiliation{e-trees.Japan, Inc., Daiwaunyu Building 2F, 2-9-2 Owadamachi, Hachioji, Tokyo 192-0045, Japan}

\author{T.\,Miyoshi}
\affiliation{Center for Quantum Information and Quantum Biology, \tcolorOrange{The University of Osaka}, 1-2 Machikaneyama, Toyonaka, Osaka, 560-0043, Japan}
\affiliation{\tcolorOrange{QuEL, Inc., Hachioji ON Building 5F, 4-7-14 Myojincho, Hachioji, Tokyo, Japan}}
\affiliation{e-trees.Japan, Inc., Daiwaunyu Building 2F, 2-9-2 Owadamachi, Hachioji, Tokyo 192-0045, Japan}

\author{Y.\,Kurimoto}
\affiliation{\tcolorOrange{QuEL, Inc., Hachioji ON Building 5F, 4-7-14 Myojincho, Hachioji, Tokyo, Japan}}

\author{Y.\,Sugita}
\affiliation{\tcolorOrange{QuEL, Inc., Hachioji ON Building 5F, 4-7-14 Myojincho, Hachioji, Tokyo, Japan}}

\author{Y.\,Ito}
\affiliation{\tcolorOrange{QuEL, Inc., Hachioji ON Building 5F, 4-7-14 Myojincho, Hachioji, Tokyo, Japan}}

\author{Y.\,Suzuki}
\affiliation{NTT Computer and Data Science Laboratories, NTT Corporation, Musashino 180-8585, Japan}

\author{P.\,A.\,Spring}
\affiliation{RIKEN Center for Quantum Computing, Wako, Saitama 351-0198, Japan}

\author{S.\,Wang}
\affiliation{RIKEN Center for Quantum Computing, Wako, Saitama 351-0198, Japan}

\author{S.\,Tamate}
\affiliation{RIKEN Center for Quantum Computing, Wako, Saitama 351-0198, Japan}

\author{Y.\,Tabuchi}
\affiliation{RIKEN Center for Quantum Computing, Wako, Saitama 351-0198, Japan}

\author{Y.\,Nakamura}
\affiliation{RIKEN Center for Quantum Computing, Wako, Saitama 351-0198, Japan}
\affiliation{Department of Applied Physics, Graduate School of Engineering, \tcolorOrange{The} University of Tokyo, 7-3-1 Hongo, Bunkyo-ku, Tokyo 113-8656, Japan}

\author{K.\,Ogawa}
\affiliation{Center for Quantum Information and Quantum Biology, \tcolorOrange{The University of Osaka}, 1-2 Machikaneyama, Toyonaka, Osaka, 560-0043, Japan} 

\author{M.\,Negoro}
\affiliation{Center for Quantum Information and Quantum Biology, \tcolorOrange{The University of Osaka}, 1-2 Machikaneyama, Toyonaka, Osaka, 560-0043, Japan}
\affiliation{\tcolorOrange{QuEL, Inc., Hachioji ON Building 5F, 4-7-14 Myojincho, Hachioji, Tokyo, Japan}}

\date{\today}

\begin{abstract}
In conventional architectures of superconducting quantum computers, each qubit is connected to its own control line, leading to a commensurate increase in the number of microwave lines as the system scales. 
Frequency-multiplexed \tcolorOrange{qubit control} addresses this problem by enabling multiple qubits to share a single microwave line. 
However, it can cause unwanted excitation of non-target qubits, especially when the detuning between qubits is smaller than the pulse bandwidth. 
Here, we propose a selective-excitation-pulse (SEP) technique that suppresses unwanted excitations by shaping a drive pulse to create null points at non-target qubit frequencies.
In a proof-of-concept experiment with three fixed-frequency transmon qubits, we demonstrate that the SEP technique achieves single-qubit gate fidelities comparable to those obtained with conventional Gaussian pulses while effectively suppressing unwanted excitations in non-target qubits.
These results highlight the SEP technique as a promising tool for enhancing frequency-multiplexed \tcolorOrange{qubit control}.
\end{abstract}
\maketitle

\begin{figure}[t]
\centering
\includegraphics[width=3.375in]{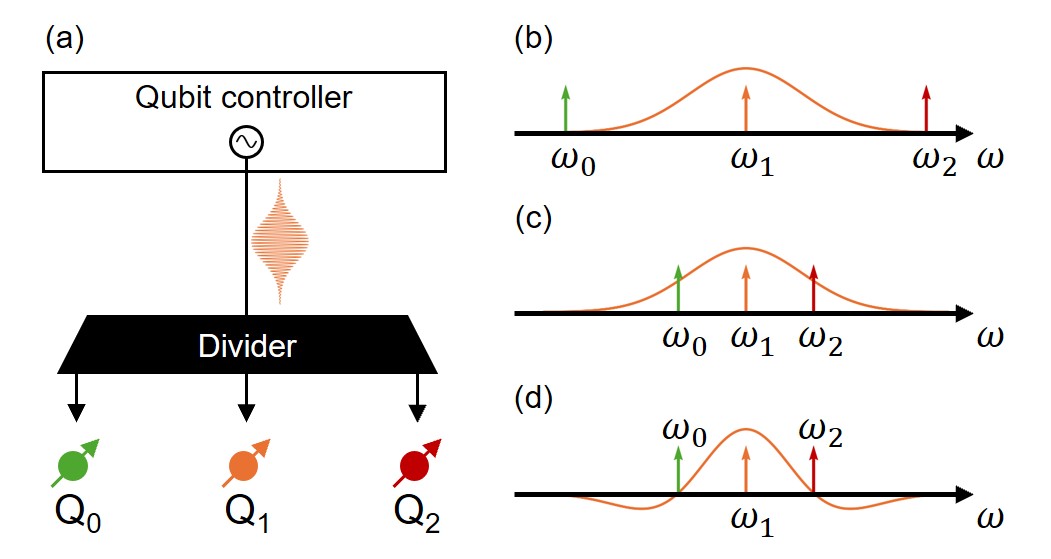}
\caption{\label{Fig: Introduction}{
(a)~Frequency-multiplexed qubit-control architecture. Multiple qubits can be manipulated using a single microwave line. 
(b)~\tcolorOrange{Qubit-frequency spacing} wider than the bandwidth of the drive pulse\tcolorOrange{.} \tcolorOrange{The off-resonant} excitation is mitigated.
(c)~\tcolorOrange{Narrow qubit-frequency spacing.} \tcolorOrange{The} bandwidth of the drive pulse \tcolorOrange{overlaps with} the transition frequencies of non-target qubits.
(d)~Selective excitation pulse (SEP). 
The frequency profile of the SEP exhibits null points in the frequency-domain amplitude at the desired frequency. 
Aligning these null points with the frequencies of non-target qubits effectively mitigates unwanted excitations.
}
}
\end{figure}

A notable challenge in scaling up superconducting quantum processors is \tcolorOrange{the wiring} scheme, which imposes severe space constraints within \tcolorOrange{a refrigerator} and generates passive heat loads that can exceed the cooling capacity when the number of qubits is high~\cite{Krinner2019-ro}.
Frequency-multiplexed \tcolorOrange{qubit control} reduces the number of control lines needed for qubit manipulation~\cite{Huang2022-va, Takeuchi2023-om, Takeuchi2024-vd, Shi2023-fd, Zhao2023-ku, Van_Dijk2019-zd, Van_Dijk2020-lq, Ohira2024-es}.
In this architecture, a single control line connects the electronics at room temperature to multiple qubits in the cryogenic environment.
Inside the dilution refrigerator, the microwave signal is distributed to individual qubits through a divider, as illustrated in Fig.~\ref{Fig: Introduction}(a).
As schematically shown in Fig.~\ref{Fig: Introduction}(b), by allocating qubit frequencies to ensure sufficient detuning relative to achievable Rabi frequencies~\cite{Van_Dijk2019-zd, Mills2022-rv, Lawrie2023-by}, it is possible to suppress unwanted \tcolorOrange{excitations of non-target qubits}.

However, this signal distribution can induce off-resonant excitations \tcolorOrange{of} non-target \tcolorOrange{qubits,} when the detuning is narrower than the pulse bandwidth [Fig.~\ref{Fig: Introduction}(c)], degrading the fidelity of quantum gate operations~\cite{Vandersypen2005-ba}.
Typically, for superconducting qubits, pulse lengths of tens of nanoseconds are used for single-qubit gate operations, resulting in a drive pulse bandwidth in the tens of megahertz range. 
The bandwidth of typical qubit controllers is around a few gigahertz~\cite{Van_Dijk2020-lq}, limiting the multiplexing factor to tens of qubits per control line. 
At this level of multiplexing, realizing a fault-tolerant quantum processor with millions of qubits would still require approximately~$\mathcal{O}(10^{5})$ control lines. 
This already exceeds the physical limitations of the dilution refrigerator~\cite{Krinner2019-ro}, implying that the degree of multiplexing is still insufficient.

In this work, we propose a selective-excitation-pulse~(SEP) technique to enhance the multiplexing factor of the frequency-multiplexed qubit-control architecture. 
The SEP technique involves designing a drive pulse that exhibits a frequency profile~\tcolorOrange{$S(\omega)$} with null points at the frequencies of non-target qubits~\cite{Linden1999-pv, Vandersypen2005-ba, Motzoi2009-ng, Chen2016-lm, Hyyppa2024-kp, Yi2024-pw, Yang2024-my}.
Therefore, the SEP technique enables the selective excitation of target qubits that share a single microwave line, even when the qubit frequencies are densely allocated, as illustrated in Fig.~\ref{Fig: Introduction}(d).

By applying the SEP technique, we experimentally demonstrate the selective excitation of a single qubit out of three using a shared microwave line.
We implement single-qubit gates based on the SEP technique and evaluate their performance using randomized benchmarking~\cite{Magesan2011-rx}.
The average gate fidelity confirms the effectiveness of the SEP technique.
Integrating the SEP technique with the frequency-multiplexed qubit-control architecture enhances the ability to simultaneously control a larger number of qubits using fewer microwave lines.
\reviseAdd{
We note that, while there is certainly potential for extending the SEP technique to frequency-multiplexed qubit control under conditions of frequency separations comparable to the Rabi frequency, this presents important challenges that warrant further investigation and lie beyond the scope of the present study.
In particular, the calibration overhead may increase significantly with the number of qubits sharing a control line.
}

We now outline the general concept of the SEP technique.
Suppose a system comprises~$N$ qubits connected to a single microwave line via a divider, as illustrated in \tcolorOrange{Fig.~\ref{Fig: Introduction}(a)}.
In this setup, when the qubits are driven by a microwave pulse at frequency~$\omega_{\textrm{d}}$, the Hamiltonian in the rotating frame of the drive frequency is given by
\begin{equation}
    \begin{aligned}[b]
    \hat{H} =& -\frac{1}{2}
    \sum_{j}{
        \Delta_{j}\hat{\sigma}_{j,z}
    } + \frac{1}{2}
    \sum_{j}{
    \left[ s_{x}(t) \hat{\sigma}_{j,x} - s_{y}(t) \hat{\sigma}_{j,y}\right]
    }, \label{Eq: Effective Hamiltonian(pauli)}
    \end{aligned}
\end{equation}
where~$\Delta_{j} = \omega_{j}-\omega_{\textrm{d}}$ represents the detuning between the drive frequency~$\omega_{\textrm{d}}$ and the resonance frequency of the \tcolorOrange{$j$th} qubit~$\omega_{j}$. Here,~$\hbar$ is set to 1.
The operators~$\hat{\sigma}_{j,x},~\hat{\sigma}_{j,y}$ and~$\hat{\sigma}_{j,z}$ are the Pauli operators for the \tcolorOrange{$j$th} qubit. 
Additionally,~$s_{x}(t)$ and~$s_{y}(t)$ are the real and imaginary parts of the waveform~$s(t)$, respectively.
When the detuning~$\Delta_{j}$ is small, a drive pulse intended for a target qubit \tcolorOrange{may} excite non-target qubits [Fig.~\ref{Fig: Introduction}(c)].

The essence of the SEP technique involves designing a pulse that exhibits null points at the frequencies of non-target qubits, thereby preventing their unintended excitation.
Let~$\mathcal{Q}_{\textrm{NT}}$ denote the set of non-target qubits. The frequency profile of the SEP is described by the following equation:
\begin{equation}
    \begin{aligned}
    S_{\mathrm{SEP}}(\omega) = A\left[\prod_{j \in \mathcal{Q}_{\textrm{NT}}}{(\omega-\omega_{j})}\right]\exp\left[-\frac{(\omega-\omega_{d})^2}{2\sigma^2}\right],
    \end{aligned}
    \label{Eq: Frequency Profile}
\end{equation}
where~$A$ is the amplitude and~$\sigma$ determines the width of the Gaussian envelope. 
The frequency of the target qubit is excluded from the product to ensure that the frequency profile does not have a null point at the target qubit frequency.

\begin{figure}[t]
    \includegraphics[width=8.5cm]{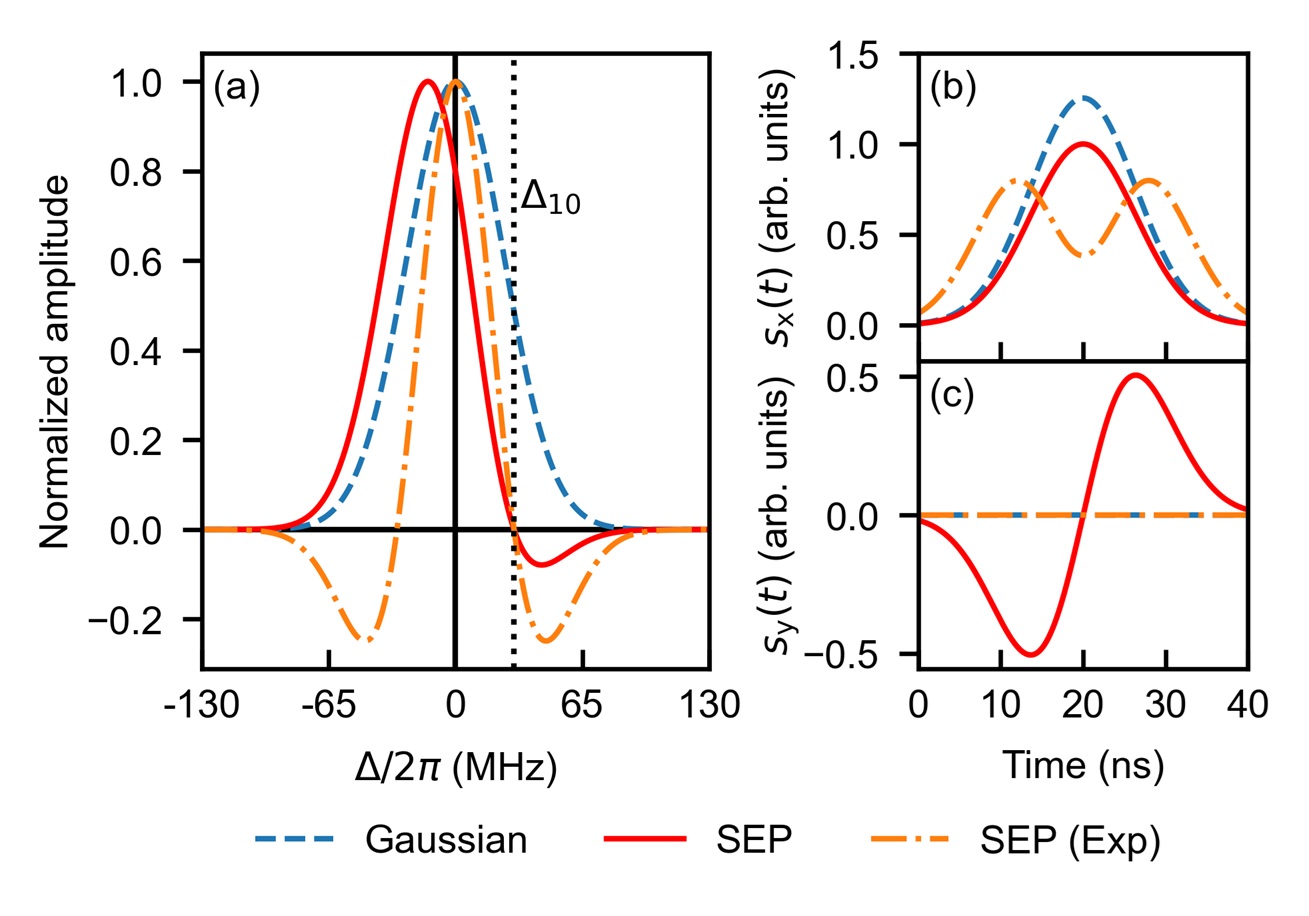}
    \caption{
    (a)~Frequency profiles of the control pulses. The dashed curve shows the Gaussian profile, the solid curve represents \tcolorOrange{the profile~$S_{\mathrm{SEP}}(\omega)$}, and the dot-dashed curve corresponds to the profile~\tcolorOrange{$S_{\mathrm{Exp}}(\omega)$}. Detailed parameters used to plot the frequency profiles are provided in the main text. 
    (b)~Real part~$s_{x}$ and~(c)~imaginary part~$s_{y}$ of the waveforms obtained from the frequency profiles shown in~(a).
    }
    \label{Fig: Pulseshaping} 
\end{figure}
\begin{table}[t]
    \centering
    \caption{Summary of the qubit and readout parameters.
    }
    \begin{tabular}{cwc{1cm}wc{1cm}wc{1cm}}
        \hline\hline
          &  Q$_{0}$ & Q$_{1}$ & Q$_{2}$\\
         \hline 
         Qubit frequency (GHz)& $8.895$ & $8.818$ & $8.792$\\ 
         Anharmonicity (MHz)  &$-411$  & $-433$ & $-413$ \\
         \tcolorOrange{Energy-relaxation} time ($\rm{\mu}$s) & $25$ & $18$& $18$  \\
         Echo dephasing time ($\rm{\mu}$s) & $21$ & $17$ & $18$\\  
         Readout-resonator frequency (GHz) & $10.456$ & $10.466$ & $10.518$ \\
         \hline\hline
    \end{tabular}
    \label{Tab: System information}
\end{table}

In the experiment, we employ a slightly modified frequency profile given as follows.
\begin{equation}
   \begin{aligned}
   S_{\mathrm{Exp}}(\omega) = A \left[\prod_{j \in \mathcal{Q}_{\textrm{NT}}}{(\omega-\omega_{j})(\omega-2\omega_{\textrm{d}}+\omega_{j})}\right] \\
   \times \exp\left[-\frac{(\omega-\omega_{\textrm{d}})^2}{2\sigma^2}\right].
   \end{aligned}
   \label{Eq: Frequency Profile (Experiment)}
\end{equation}
The correction transforms the frequency profile to be symmetric about the drive frequency~$\omega_\textrm{d}$, helping to minimize phase shifts due to \tcolorOrange{ac Stark} shifts~\cite{Chen2016-lm}.

\begin{figure*}[t]
    \centering
    \includegraphics[width=17.0cm]{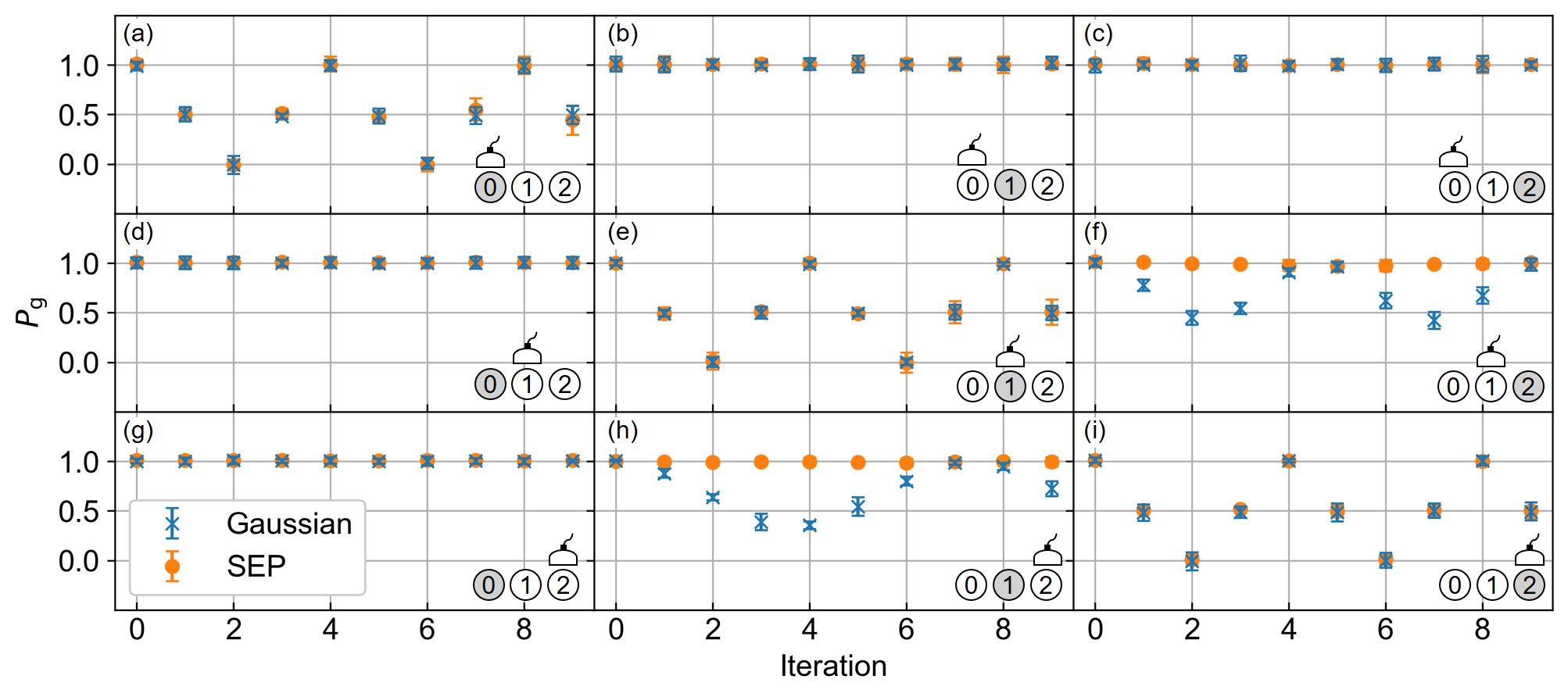}
    \caption{
    Rabi oscillations observed when applying repeated~$X_{\pi/2}$ gates using Gaussian pulses (crosses) and the SEPs (dots).
    (a)--(c)~Measured results for~Q$_{0}$, (d)--(f)~for~Q$_{1}$, and (g)--(i)~for~Q$_{2}$.
    Each data point represents the average of five experiments, with error bars indicating 1$\sigma$ confidence intervals.
    \tcolorOrange{In insets, the} target qubit is shown in \tcolorOrange{gray,} and the non-target qubits are shown in white.
    The detector icon indicates the qubit that was measured to obtain the results.
    }
    \label{Fig: Rabi}
\end{figure*}

Now, let us consider a concrete example with two qubits,~Q$_0$ and~Q$_1$, designated as the target qubit and the non-target qubit, respectively, with frequencies~$\omega_0$ and~$\omega_1$.
The frequency profile of the SEP, designed to suppress excitation in~Q$_1$ is shown in Fig.~\ref{Fig: Pulseshaping}(a).
Note that the corresponding frequency profiles are plotted as a function of the detuning~$\Delta$ relative to the frequency of~Q$_0$.
In Fig.~\ref{Fig: Pulseshaping}(a), the solid and the dot-dashed curves represent the frequency profiles based on \tcolorOrange{Eqs.~(\ref{Eq: Frequency Profile}) and~(\ref{Eq: Frequency Profile (Experiment)})}, respectively, with parameters~$\sigma/2\pi=25$~MHz and~$\tcolorOrange{\Delta_{10}}/2\pi=30$~MHz.
Here,~$\tcolorOrange{\Delta_{10}}=\omega_1-\omega_0$ represents the frequency difference between~Q$_0$ and~Q$_1$.
Additionally, the frequency profile of a Gaussian pulse is shown as the dashed curve, plotted using Eq.~(\ref{Eq: Frequency Profile}) with the non-target qubit terms set to 1, with a parameter~$\sigma/2\pi=25$~MHz.
Note that, for all frequency profiles shown in Fig.~\ref{Fig: Pulseshaping}(a), the amplitude~$A$ is set such that the peak of each frequency profile reaches \tcolorOrange{unity}.

The corresponding time-domain signal~$s(t)$ is then obtained by applying the inverse Fourier transform to the frequency profile:
\begin{equation}
    \begin{aligned}
    s(t) = \mathcal{F}^{-1} [ S(\omega) ].
    \end{aligned} \label{Eq: Inverse Fourier transform}
\end{equation}
The real and imaginary parts of the time-domain signal, obtained from the frequency profiles shown in Fig.~\ref{Fig: Pulseshaping}(a), are shown in Figs.~\ref{Fig: Pulseshaping}(b) and (c), respectively.
By using these signals to drive the target qubit, \tcolorOrange{we suppress excitations} in non-target \tcolorOrange{qubits~\cite{Motzoi2013-gs}.}

\reviseAdd{
The SEP technique exhibits discontinuities at the initial and final points of the waveform, as shown in Fig.~\ref{Fig: Pulseshaping}(b).
These discontinuities can induce spectral leakage and associated control errors. 
To suppress such effects, the smoothness of the waveform is improved by appropriately tuning the spectral width parameter~$\sigma$ in Eq.~(\ref{Eq: Frequency Profile}). 
A detailed discussion of the origin and mitigation of these discontinuities is provided in the Supplementary Material~\cite{Supplemental}.
}

To investigate the effectiveness of the SEP technique, we conducted a series of experiments.
The experimental setup employed in this study is illustrated in the Supplemental Material~\cite{Supplemental}.
All experiments were conducted using a device consisting of 64 fixed-frequency transmon qubits~\cite{Koch2007-xs, Tamate2022-ep, Watanabe2024-mv, Negoro2024-hl}. 
Our objective was to demonstrate that the SEP technique enables selective excitation even when qubits with closely spaced resonance frequencies are connected to the same microwave line. 
For this purpose, we required a set of qubits with resonance frequencies within the bandwidth of the drive pulse used in this study. 
Based on this criterion, we selected three transmon qubits, labeled~Q$_{0}$,~Q$_{1}$, and~Q$_{2}$, respectively.
The selected qubits are not nearest neighbors and thus do not have direct couplings between them.
The detailed properties of the qubits and their associated resonators used in the experiments are summarized in Table~\ref{Tab: System information}.

To experimentally demonstrate our proposed technique, we controlled the selected three qubits using a shared microwave line.
The microwave signal for qubit control was split into three separate control lines using a power divider, with each line directed to an individual qubit~\cite{Supplemental}.
While the primary motivation of this study is to reduce the number of microwave lines inside \tcolorOrange{the dilution} refrigerator by installing \tcolorOrange{a divider to} split the signal just before reaching the qubits, the current experimental setup is sufficient for a proof-of-concept demonstration.

\begin{figure*}[t]
    \centering
    \includegraphics[width=17.0cm]{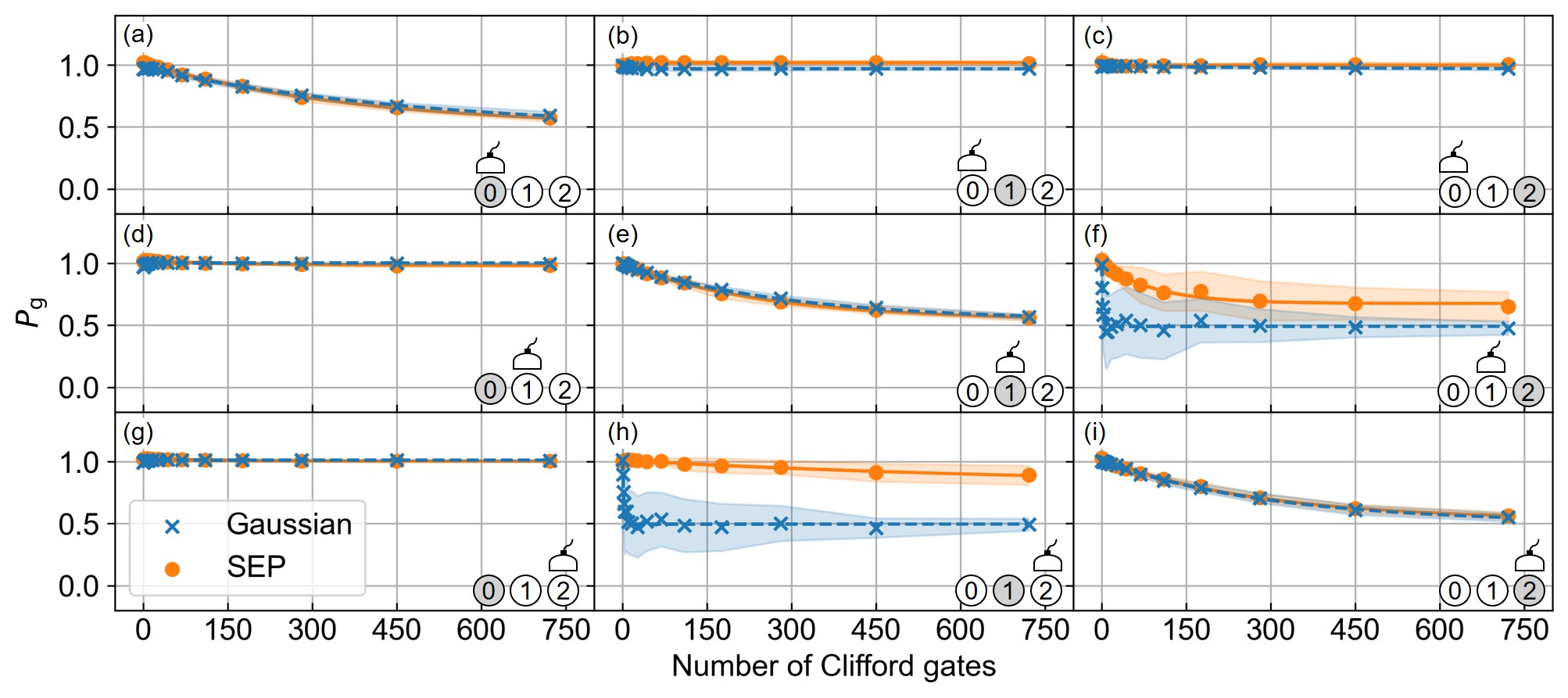}
    \caption{
    Randomized benchmarking of single-qubit gates for Gaussian pulses (crosses) and the SEPs (dots).
    (a)--(c)~Measured results for~Q$_{0}$, (d)--(f)~for~Q$_{1}$, and (g)--(i)~for~Q$_{2}$.
    Data points represent the average of 50 different random Clifford sequences, with shaded regions indicating the standard deviation across the 50 results for each Clifford-gate number.
    The~$X_{\pi/2}$ gate was calibrated in the SEP technique.
    The SEP technique achieved an average gate fidelity comparable to that of Gaussian pulses.
    }
    \label{Fig: RB}
\end{figure*}

In the first experiment, we implemented an~$X_{\pi/2}$ gate using the SEP technique and a Gaussian pulse with a duration of 40~ns to assess the suppression of unwanted excitations in non-target qubits.
We calibrated the amplitude~$A$ of the frequency \tcolorOrange{profile, given} in \tcolorOrange{Eq.~(\ref{Eq: Frequency Profile (Experiment)}) to} achieve a rotation angle of~$\pi/2$.
To compensate for phase drift caused by \tcolorOrange{ac Stark} shifts during gate operations, we introduced and fine-tuned an additional detuning parameter~$\delta$ of the frequency profile~\cite{Supplemental}. 
The waveforms used to implement the~$X_{\pi/2}$ gate and \tcolorOrange{their} parameters are provided in the Supplemental Material.

Using the SEP technique and Gaussian pulses, we repeatedly applied the~$X_{\pi/2}$ \tcolorOrange{gates} to the three qubits and measured the population of each qubit. 
We executed 1,000 experiments to obtain \tcolorOrange{$P_{\textrm{g}}$}, the probability of finding the qubit in its ground state.
After each measurement, we waited for a period of at least~$10~T_{1}$ to ensure that the qubit state was properly initialized.

In Fig.~\ref{Fig: Rabi}, we present the Rabi oscillations \tcolorOrange{under} repeated~$X_{\pi/2}$ gates. 
As shown in Figs.~\ref{Fig: Rabi}(a), (e) and~(i), the~$X_{\pi/2}$ gate is properly calibrated using both the \tcolorOrange{SEP and} Gaussian pulses.
\tcolorOrange{Off-resonant} excitations are observed in the non-target qubits when Gaussian pulses are used, as illustrated in Figs.~\ref{Fig: Rabi}(f) and~(h). 
In contrast, the SEP technique effectively suppresses these unwanted excitations.

Next, to evaluate the average fidelity of the single-qubit gate using the SEP technique, we performed single-qubit randomized benchmarking~\cite{Magesan2011-rx}. 
All single-qubit Clifford gates were composed of~$X_{\pi/2}$ gates and virtual-Z gates~\cite{McKay2017-sh}.
For comparison, we also evaluated the single-qubit gate fidelity using Gaussian pulses.
Figures~\ref{Fig: RB}(a), (e) and~(i) present the results of the single-qubit randomized benchmarking. 
Using the single-qubit gate with the SEP technique, the average single-qubit Clifford gate fidelities were determined to be~99.87(1)\%, 99.80(2)\%, and~99.82(2)\% for target qubits ~Q$_{0}$,~Q$_{1}$, and~Q$_{2}$, respectively.
In comparison, the Gaussian pulses yielded fidelities of~99.90(2)\%, 99.83(2)\%, and~99.83(1)\% for~Q$_{0}$,~Q$_{1}$, and~Q$_{2}$, respectively.
The fidelities achieved with the SEP technique are thus comparable to those obtained with Gaussian pulses.

In addition to the gate fidelity, we examined the excitation probability for non-target qubits to quantify unwanted excitations. 
We fit the populations~\tcolorOrange{$P_{\textrm{g}}$} to a decay model~$Ap^{L}+B$, where~$p$ is the depolarizing rate.
Here, the coefficients~$A$ and~$B$ account for state preparation errors and the asymptotic population value as the sequence length increases.

The obtained results for non-target qubits are shown in Figs.~\ref{Fig: RB}(f) and~(h). 
There is a clear difference in the excitation levels between experiments using the SEP technique and those using Gaussian pulses.
To quantify this difference in off-resonant excitation between the SEP technique and Gaussian pulses, we introduce the excitation rate, defined as~$\Gamma_{\textrm{ex}} \equiv B(1-p)/2$.
Physically, this rate reflects the balance between state excitation and energy relaxation over the sequence length.

Based on the results shown in Fig.~\ref{Fig: RB}(f)[(h)], the excitation rate is~$0.2$\%($0.01$\%) for the SEP technique and~$10$\%($7$\%) for Gaussian pulses.
These results indicate that using the SEP technique effectively suppresses excitations compared to Gaussian pulses.
\reviseAdd{
Although Gaussian pulses are used as the base pulse in this study, the SEP technique is compatible with other basis-pulse shapes, including raised-cosine pulses.
Additionally, SEP may be compatible with techniques such as the derivative removal by adiabatic gate~(DRAG)~\cite{Motzoi2013-gs, Wang2025-xq}.
\reviseAdd{We numerically explored this possibility and found that the combination can suppress leakage to the order of~$10^{-4}$ with parameters: a pulse duration~$T=16\,\mathrm{ns}$, the detuning between the target qubit and the non-target qubit~$46$~MHz, and an anharmonicity~$\alpha/2\pi=-300\,\mathrm{MHz}$, potentially enabling faster and more robust gate implementations.
}
}

In this work, we demonstrated the selective control of three transmon qubits through a single microwave line by leveraging the selective-excitation-pulse~(SEP) technique. 
We achieved gate fidelities comparable to Gaussian-pulse-based operations while significantly suppressing unwanted excitations in non-target qubits. 
Our results suggest that the SEP technique can serve as a valuable tool for implementing efficient, scalable qubit control in advanced quantum computing architectures.

\begin{acknowledgments}
This research was supported by JST COI-NEXT (Grant No. JPMJPF2014), JST Moonshot R$\&$D (Grant No. JPMJMS2067, Grant No. JPMJMS226A), MEXT  Q-LEAP  (Grant  No.  JPMXS0118068682), JST PRESTO (Grant No. JPMJPR23F2) and JST the establishment of university fellowships towards the creation of science technology innovation (Grant No. JPMJFS2125).
\end{acknowledgments}

\nocite{Chen2016-lm}
\nocite{Sumida2024-ci}
\nocite{Blais2004-kh}

\bibliographystyle{apsrev4-2}
\bibliography{ref}

\pagebreak
\widetext

\beginsupplement

\begin{center}
\textbf{\large Supplemental Material for ``Selective Excitation of Superconducting Qubits with~a~Shared Control Line through Pulse Shaping''}
\\~\\
R.~Matsuda,$^{1,2,\ast}$ R.~Ohira,$^{3,\ast}$ T.~Sumida,$^{3}$ H.~Shiomi,$^{2,3}$ A.~Machino,$^{1,2}$ S.~Morisaka,$^{2,3}$ \\
K.~Koike,$^{4}$ T.~Miyoshi,$^{2,3,4}$ Y.~Kurimoto,$^{3}$ Y.~Sugita,$^{3}$ Y.~Ito,$^{3}$ Y.~Suzuki,$^{5}$ P.~A.~Spring,$^{6}$ \\
S.~Wang,$^{6}$, S.~Tamate,$^{6}$ Y.~Tabuchi,$^{6}$ Y.~Nakamura,$^{6,7}$ K.~Ogawa,$^{2}$ M.~Negoro$^{2,3}$
\\~\\
\small{
$^{1}$\textit{Graduate School of Engineering Science, \tcolorOrange{The University of Osaka}, \\ 1-3 Machikaneyama, Toyonaka, Osaka, 560-8531, Japan} \\
$^{2}$\textit{Center for Quantum Information and Quantum Biology, \tcolorOrange{The University of Osaka}, \\ 1-2 Machikaneyama, Toyonaka, Osaka, 560-0043, Japan} \\
$^{3}$\textit{QuEL, Inc., Hachioji ON Building 5F, 4-7-14 Myojincho, Hachioji, Tokyo, Japan} \\
$^{4}$\textit{e-trees.Japan, Inc., Daiwaunyu Building 2F, 2-9-2 Owadamachi, Hachioji, Tokyo 192-0045, Japan} \\
$^{5}$\textit{NTT Computer and Data Science Laboratories, NTT Corporation, Musashino 180-8585, Japan} \\
$^{6}$\textit{RIKEN Center for Quantum Computing, Wako, Saitama 351-0198, Japan} \\
$^{7}$\textit{Department of Applied Physics, Graduate School of Engineering, \\ \tcolorOrange{The} University of Tokyo, 7-3-1 Hongo, Bunkyo-ku, Tokyo 113-8656, Japan} \\
}
\end{center}

\onecolumngrid

\section{\label{Sec in suppl: SEP}
Suppression of Unintended Excitation Using Selective Excitation Pulse
}

\reviseAdd{
To illustrate how a drive pulse based on the SEP technique suppresses unintended excitation, we analyze the dynamics of a non-target qubit under the Hamiltonian given by
\begin{equation}
    \begin{aligned}[b]
    \hat{H} =& 
    \sum_{j}{
        \Delta_{j}\hat{a}^{\dagger}_{j}\hat{a}_{j}
    } + \frac{1}{2}
    \sum_{j}{
    \left[ s(t) \hat{a}^{\dagger}_{j} + s^{\ast}(t) \hat{a}_{j}\right]
    }, \label{Eq in suppl: Effective Hamiltonian(pauli)}
    \end{aligned}
\end{equation}
where~$\Delta_{j} = \omega_{j} - \omega_{\textrm{d}}$ represents the detuning between the drive frequency $\omega_{\textrm{d}}$ and the resonance frequency of the $j$-th qubit~$\omega_{j}$.
$\hat{a}^{\dagger}_{j}\,(\hat{a}_{j})$ denotes the creation~(annihilation) operator of the~$j$-th qubit.
$s(t)$~indicates an envelope of pulse used to drive qubits. Note that we set $\hbar=1$.}

\reviseAdd{
Here, we consider a situation with two qubits: the target qubit~Q$_0$ and the non-target qubit~Q$_1$ with respective frequencies $\omega_0$ and $\omega_1$. Both qubits are initially prepared in the ground state~$\ket{\mathrm{g}}$. 
Although our analysis focuses on a two-qubit system for clarity, the SEP technique and its suppression of unintended excitations can be generalized to systems with~$n$ qubits.
}

\reviseAdd{
Using the Hamiltonian and the three types of waveforms, we calculate~$P_{\textrm{e}}$, the probability of finding the non-target qubit in the first excited state~$\ket{\textrm{e}}$ after a pulse duration of~$T$. 
The first waveform is a Gaussian pulse, while the second and third waveforms are derived from the following frequency profiles:
\begin{equation}
    \begin{aligned}
    S_{\mathrm{asym}}(\omega) = A(\omega-\omega_{1}) \exp\left[-\frac{(\omega-\omega_{\textrm{d}})^2}{2\sigma^2}\right],
    \end{aligned}
    \label{Eq in suppl: Frequency Profile}
\end{equation}
and
\begin{equation}
   \begin{aligned}
   S_{\mathrm{sym}}(\omega) = A (\omega-\omega_{1})(\omega-2\omega_{\textrm{d}}+\omega_{1}) \exp\left[-\frac{(\omega-\omega_{\textrm{d}})^2}{2\sigma^2}\right].
   \end{aligned}
   \label{Eq in suppl: Frequency Profile (Experiment)}
\end{equation}
Here,~$A$ is the amplitude and~$\sigma$ determines the width of the Gaussian envelope.
The parameters used for this calculation are~$\sigma/2\pi = 1/T$.
The amplitude~$A$ and the drive frequency~$\omega_{\textrm{d}}$ are chosen to satisfy~$\int_{0}^{T}s_{\mathrm{asym\,(sym)}}(t)\,dt=\pi/2$ and~$\Delta_{0} = 0$, where~$s_{\mathrm{asym\,(sym)}}(t)$ is respectively obtained by applying the inverse Fourier transform to~$S_{\mathrm{asym\,(sym)}}(\omega)$, followed by truncation within the pulse duration~$T$.
}

\reviseAdd{
Figure~\ref{Fig in suppl: Numerical results} summarizes~\(P_{\textrm{e}}\) as functions of the pulse duration~\(T\) and the detuning~\(\Delta_{1}\).
Figures~\ref{Fig in suppl: Numerical results}(a)--(c) show the excitation probability of the target qubit~Q$_0$ driven by different pulses:~(a) Gaussian pulse, (b)~SEP~$s_{\mathrm{asym}}(t)$, and~(c)~SEP~$s_{\mathrm{sym}}(t)$.
Figures~\ref{Fig in suppl: Numerical results}(d)--(f) show the corresponding excitation probability~\(P_{\textrm{e}}\) for the non-target qubit~Q$_1$.
For the target qubit~Q$_0$, the Gaussian pulse [Fig.~\ref{Fig in suppl: Numerical results}(a)] achieves~\(P_{\textrm{e}} > 0.49\) with relatively short durations (approximately 20~ns), while SEP-based pulses [Figs.~\ref{Fig in suppl: Numerical results}(b) and (c)] generally require longer durations to reach similar excitation levels.
In contrast, for the non-target qubit~Q$_1$, SEP-based pulses [Figs.~\ref{Fig in suppl: Numerical results}(e) and (f)] provide significantly better suppression of unwanted excitation than the Gaussian pulse [Fig.~\ref{Fig in suppl: Numerical results}(d)], particularly when the detuning~\(\Delta_1\) is small.
These results indicate that by utilizing the SEP technique, selective excitation of the target qubit can be achieved even when the frequencies of the target and non-target qubits are closely spaced.
\begin{figure}[t]
    \centering
    \includegraphics[width=17.0cm]{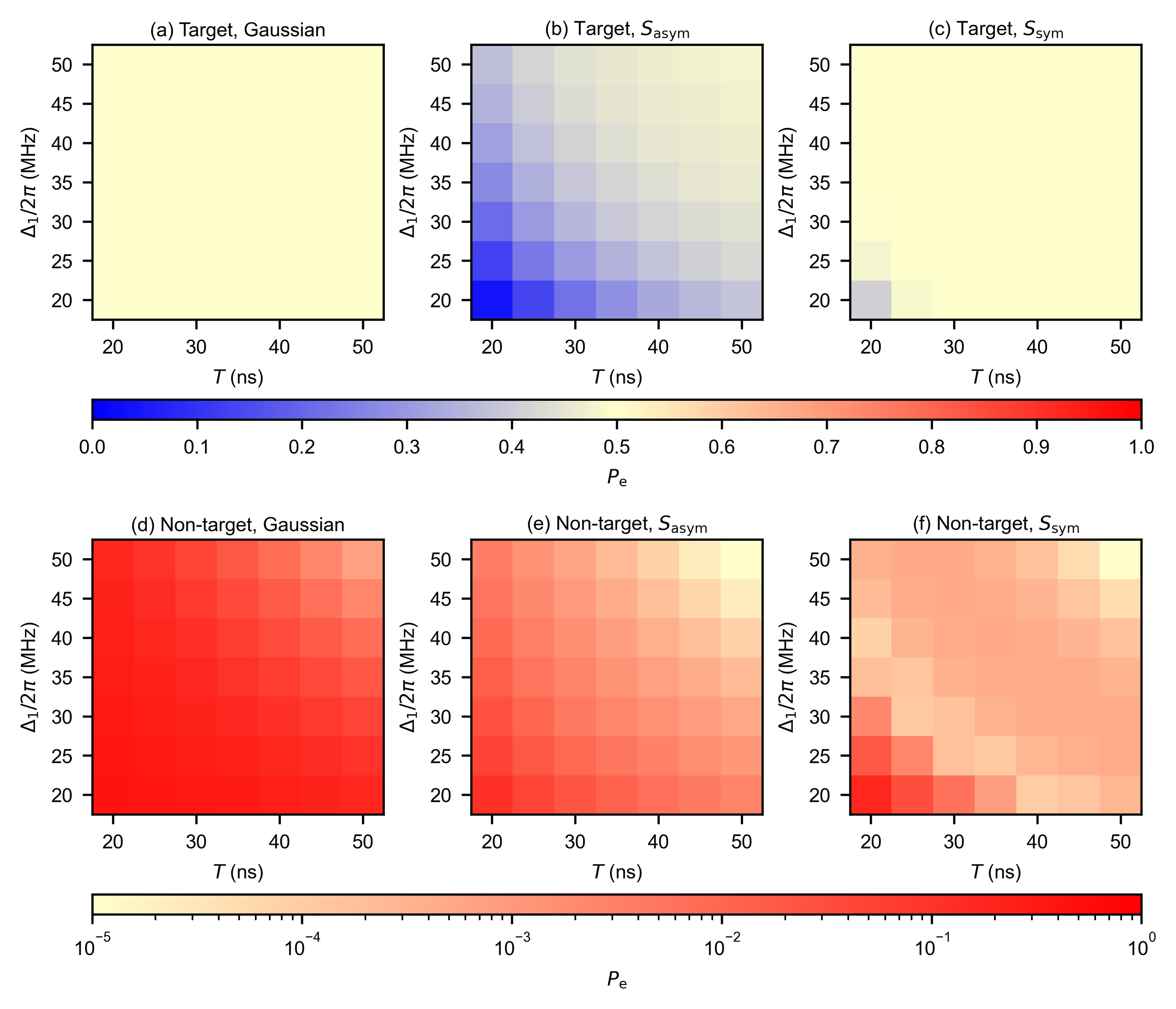}
    \caption{
    \reviseAdd{
    Numerical simulation of the excitation probability~$P_{\textrm{e}}$ of the target qubit~Q$_0$ and the non-target qubit~Q$_1$ after an~$X_{\pi/2}$ gate, shown as a function of the pulse duration~$T$ and detuning~$\Delta_1$.
    (a)--(c)~Results for~Q$_0$ using three waveform types: (a)~Gaussian, (b)~$s_{\mathrm{asym}}(t)$, and (c)~$s_{\mathrm{sym}}(t)$.
    (d)--(f)~Results for~Q$_1$, using the same waveform types as in~(a)--(c).
    }
    }
    \label{Fig in suppl: Numerical results}
\end{figure}
}

\clearpage
\section{\protect\tcolorOrange{Selective-Excitation} Pulse for Single-Qubit Gate Operation\label{Sec in suppl: Pulse Shape}}

The frequency profile for implementing an $X_{\pi/2}$ gate using the SEP technique in the experiment is given as follows:
\begin{equation}
   \begin{aligned}
   S(\omega) = A \left[\prod_{j \in \mathcal{Q}_{\textrm{NT}}}{(\omega-\omega_{j})(\omega-2\omega_{\textrm{d}}+\omega_{j})}\right] \exp\left[-\frac{(\omega-\omega_{\textrm{d}}-\delta)^2}{2\sigma^2}\right].
   \end{aligned}
   \label{Eq in suppl: Frequency Profile (Experiment)_mod}
\end{equation}
Here, the detuning parameter $\delta$ is introduced to further fine-tune the frequency profile. 
By adjusting $\delta$, we compensate for residual \tcolorOrange{ac Stark} shifts during gate operations~\cite{Chen2016-lm}.

\begin{figure}[h]
    \centering
    \includegraphics[width=15cm]{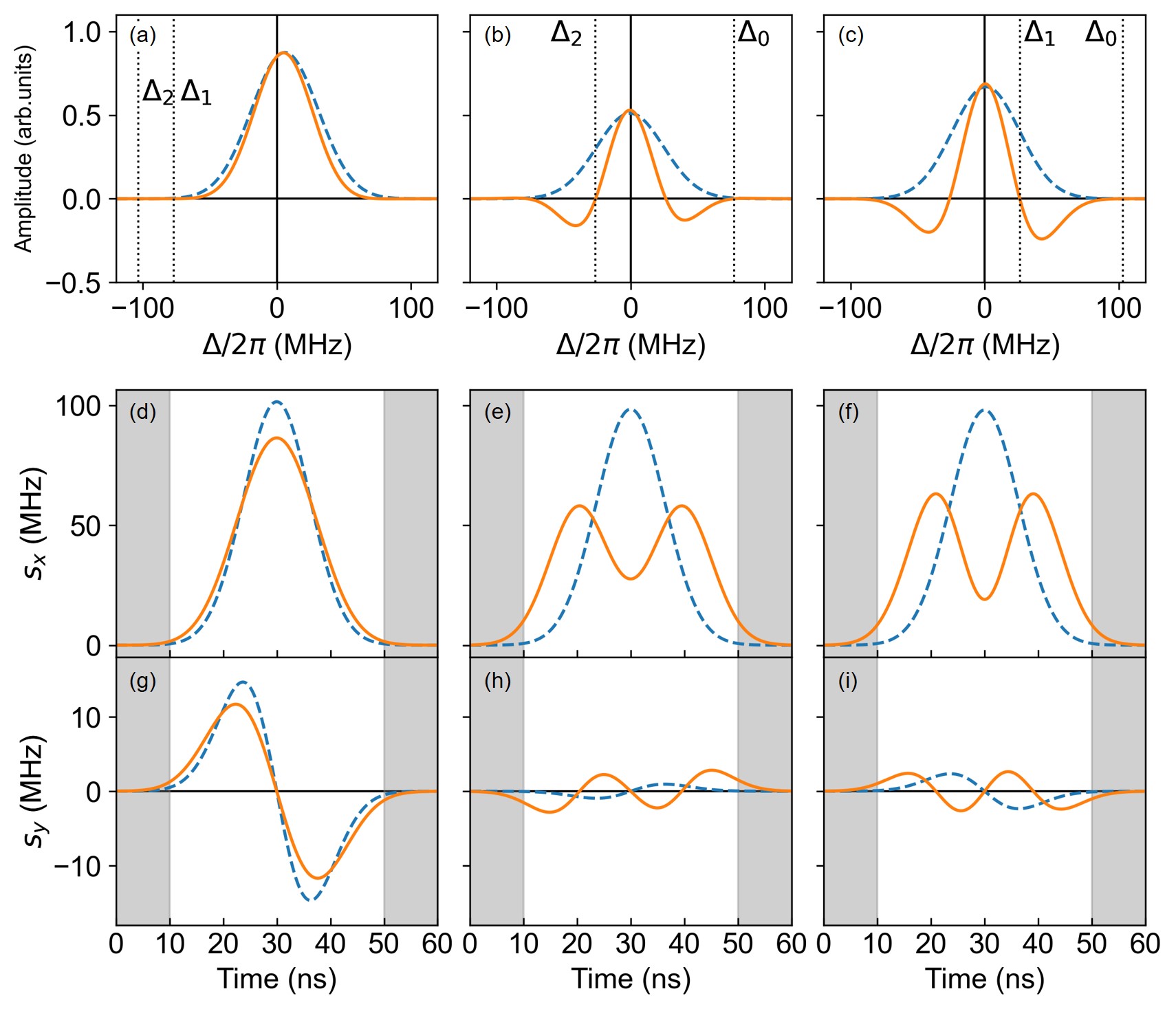}
    \caption{
    Frequency profiles and time-domain signals used to implement the~$X_{\pi/2}$ gates. 
    (a)--(c)~Frequency profiles for the target qubits~Q$_{0}$,~Q$_{1},$ and ~Q$_{2}$, respectively. The dashed and solid curves show the results for Gaussian pulses and the SEPs, respectively. 
    (d)--(i)~Time-domain signals derived from each frequency profile: (d)--(f)~real parts~$s_{x}$ and (g)--(i)~imaginary parts~$s_{y}$ of the corresponding waveforms.
    The shaded regions indicate truncated segments of the waveforms and the non-shaded regions indicate the waveforms used in the experiments.}
    \label{Fig in suppl: The waveforms used to implement gate}
\end{figure}

The waveforms corresponding to the $X_{\pi/2}$ gate are shown in Fig.~\ref{Fig in suppl: The waveforms used to implement gate}, and the detailed parameters are summarized in Table~\ref{Tab in suppl: Parameters_Gaussian_SEP}.

\begin{table}[h]
    \centering
    \caption{Parameters used \tcolorOrange{for the} Gaussian pulses and the \tcolorOrange{SEPs.}}
    \begin{tabular}{cwc{1.0cm}wc{1.0cm}wc{1.0cm}wc{1.0cm}wc{1.0cm}wc{1.0cm}}
        \hline\hline
          & \multicolumn{3}{c}{Gaussian \tcolorOrange{pulse}} & \multicolumn{3}{c}{SEP} \\
          \cline{1-7}
          \tcolorOrange{Target qubit} & Q$_{0}$ & Q$_{1}$ & Q$_{2}$ & Q$_{0}$ & Q$_{1}$ & Q$_{2}$ \\
         \hline 
         $\delta/2\pi$~(MHz)  
         & $6.019$ & $-0.3979$ & $0.9885$ 
         & $6.561$ & $-1.767$ & $1.354$ \\
         $\Delta_{j}/2\pi$~(MHz) 
         & \tcolorOrange{--} & \tcolorOrange{--} & \tcolorOrange{--} 
         & \makecell{$ -77$ \\ $ -103$} & \makecell{$ 77$ \\ $ -26$} & \makecell{$ 103$ \\ $ 26$} \\
         $\omega_{\textrm{d}}/2\pi$~(GHz) 
         & $8.895$ & $8.818$ & $8.792$ 
         & $8.895$ & $8.818$ & $8.792$ \\
         $\sigma/2\pi$~(MHz)
         & $25$ & $25$ & $25$ 
         & $25$ & $25$ & $25$ \\
         \hline\hline
    \end{tabular}
    \label{Tab in suppl: Parameters_Gaussian_SEP}
\end{table}

\clearpage
\section{Discontinuities in the Selective Excitation Pulse \label{Sec in suppl: Discontinuity}}
\reviseAdd{
Due to the finite temporal duration of the waveform, SEPs exhibit discontinuities at their start and end points, as shown in Fig.~\ref{Fig in suppl: The waveforms used to implement gate}. 
These discontinuities are expected to introduce significant spectral leakage, particularly for shorter pulses where truncation effects become more pronounced.
To mitigate this issue and improve waveform smoothness, we investigate the role of the spectral width parameter~$\sigma$ in Eq.~(\ref{Eq in suppl: Frequency Profile (Experiment)}), which defines the envelope width of the frequency-domain profile. 
The amplitude~$A$ is normalized such that the time-domain waveform~$s(t)$, obtained via inverse Fourier transform of $S(\omega)$, satisfies the area condition~$\int^{T}_{0}s(t)\,dt = \pi/2$, with a fixed truncated pulse duration~$T = 40$~ns. 
We set $\Delta_1 / 2\pi = 0.03$~GHz, where $\Delta_1$ denotes the detuning between the drive frequency~$\omega_{\mathrm{d}}$ and the non-target frequency~$\omega_1$ in Eq.~(\ref{Eq in suppl: Frequency Profile (Experiment)}).
}

\begin{figure}[h]
    \centering
    \includegraphics[width=17cm]{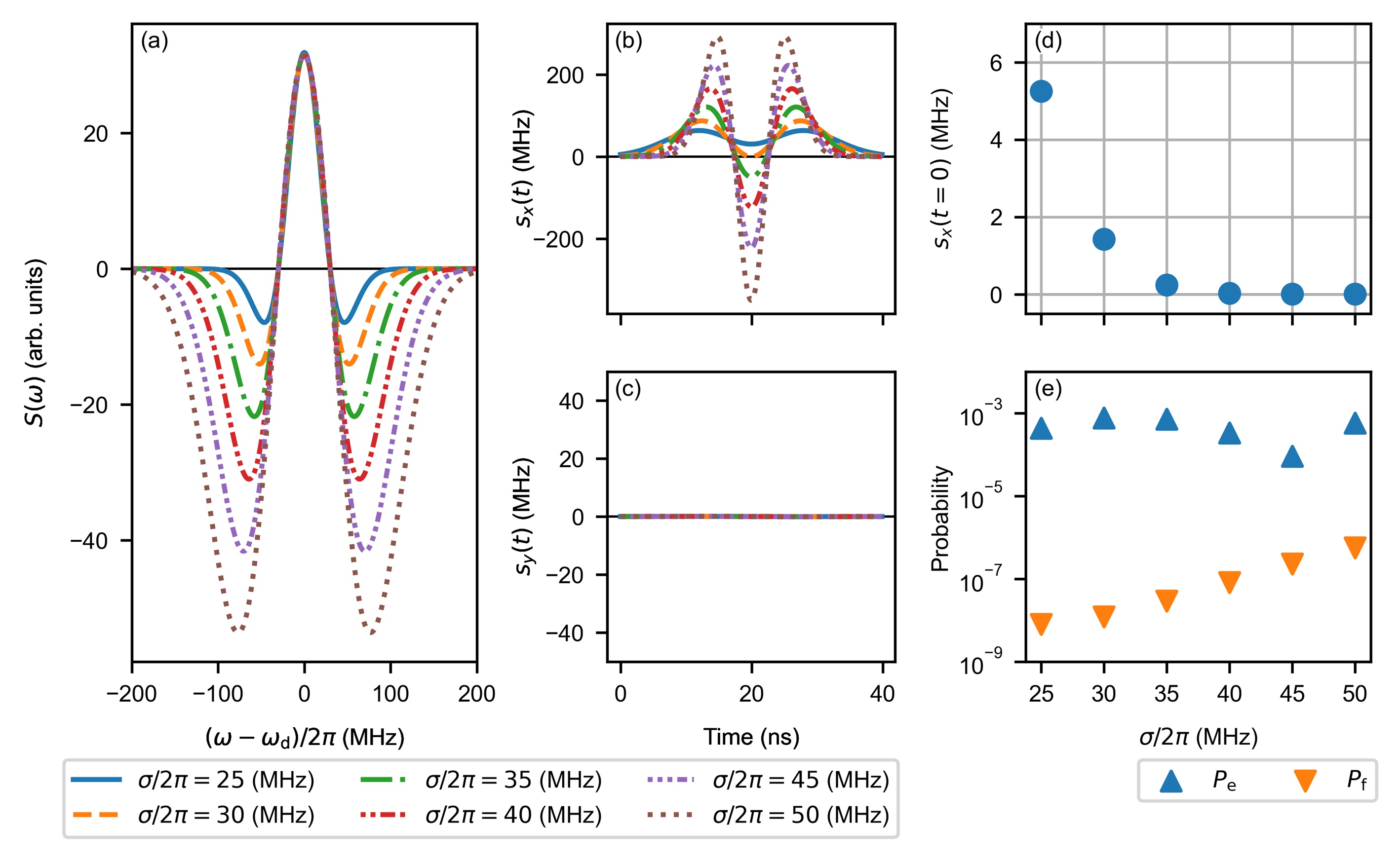}
    \caption{
    \reviseAdd{
    Analysis of the waveform discontinuity.
    (a)~Frequency profiles of the SEP given by~Eq.~(\ref{Eq in suppl: Frequency Profile (Experiment)}), with the amplitudes normalized so that~$s(t)$ satisfies the area condition $\int^{T}_{0}s(t)\,dt=\pi/2$. Here, $s(t)$ is obtained via the inverse Fourier transform of~$S(\omega)$ in~(a). 
    (b),(c)~Time-domain signals: real part~$s_{x}(t)$ and imaginary part~$s_{y}(t)$ of the waveform~$s(t)$.
    (d)~Value of~$s_x(t=0)$ as a function of~$\sigma$, showing the discontinuity due to the truncation.
    (e)~Numerical results of $P_{\textrm{e}}\,(P_\textrm{f})$, the probability of projecting the non-target qubit into the first (second) excited state $\ket{\textrm{e}}\,(\ket{\textrm{f}})$.}
    }
    \label{Fig in suppl: discontinuity}
\end{figure}

\reviseAdd{
To examine the influence of~$\sigma$, we compute the frequency-domain profiles~$S(\omega)$ and the corresponding time-domain waveforms~$s(t)$ for several values of~$\sigma/2\pi$, as shown in Figs.\ref{Fig in suppl: discontinuity}(a)--(c). 
The time-domain signals reveal that both the initial and final points of the waveform become smoother as $\sigma$ increases.
To quantify this effect, Fig.~\ref{Fig in suppl: discontinuity}(d) plots the real part~$s_x(t=0)$ as a function of~$\sigma$. 
The amplitude at the initial time point decreases as~$\sigma/2\pi$ increases, indicating the reduction in discontinuity.
Finally, for each~$s(t)$ as shown in~Figs.\ref{Fig in suppl: discontinuity}(b) and (c), we numerically determine $P_\textrm{e}\,(P_{\textrm{f}})$, the probability of projecting the non-target qubit into the first (second) excited state~$\ket{\textrm{e}}\,(\ket{\textrm{f}})$ after driven by each pulse, as a function of $\sigma$. The unwanted excitation~$P_{\textrm{e}}$ of the non-target qubit is suppressed at a level of~$10^{-4}$ for any value of $\sigma$. Although the reduction of the discontinuity is shown in Fig.~\ref{Fig in suppl: discontinuity}(e), increasing the microwave amplitude as shown in Fig.~\ref{Fig in suppl: discontinuity}(b) leads to the unintended excitation~$P_{\textrm{e}}$ and the leakage~$P_{\textrm{f}}$ of the non-target qubits.
}

\clearpage

\section{Experimental setup\label{Sec in suppl: Experimental Setup}}

\begin{figure}[h]
    \centering
    \includegraphics[width=9.0cm]{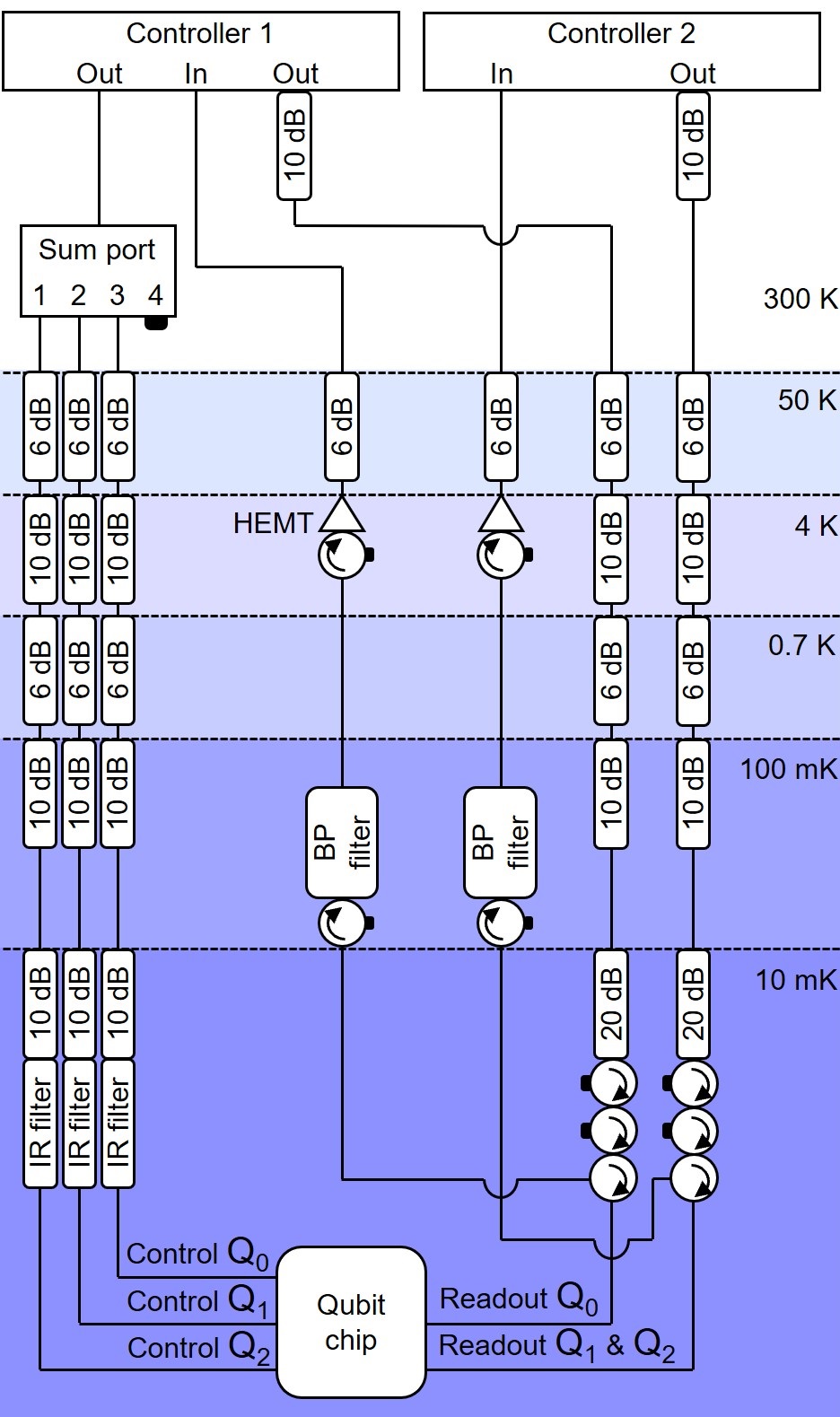}
    \caption{
    Wiring schematic \tcolorOrange{of the} experimental setup.
    Background color indicates the temperature at each stage.
    The setup allows pulses driving the target qubit to interact with non-target qubits via a signal divider. 
    }
    \label{Fig in suppl: DetailedWiring}
\end{figure}

The experimental setup employed in this study is illustrated in Fig.~\ref{Fig in suppl: DetailedWiring}.
The device consisting of 64 fixed-frequency transmon qubits was mounted to the stage of a dilution refrigerator (Bluefors XLD1000sl) at approximately 10~mK and was magnetically shielded.

All microwave signals for qubit control and readout were generated by qubit controllers (QuEL, Inc.)~\cite{Sumida2024-ci, Negoro2024-hl}.
These signals were routed to the qubits via attenuators placed at the various stages of the dilution refrigerator.

The qubit readout was performed using a dispersive readout technique~\cite{Blais2004-kh}. 
The readout resonator in the device was dispersively coupled to \tcolorOrange{one} of \tcolorOrange{the four} transmon \tcolorOrange{qubits in a unit cell}, allowing for the simultaneous readout of all four qubits. 
In this setup,~Q$_1$ and~Q$_2$ shared the same measurement line. 
The readout signals were amplified by high-electron-mobility transistor (HEMT) amplifiers located at the \tcolorOrange{4-K} stage before being routed back to the qubit controllers.

\end{document}